# Conditional maximum-entropy method for selecting prior distributions in Bayesian statistics

SUMIYOSHI ABE

*Department of Physical Engineering, Mie University, Mie 514-8507, Japan*

**Abstract**  The conditional maximum-entropy method (abbreviated here as C-MaxEnt) is formulated for selecting prior probability distributions in Bayesian statistics for parameter estimation. This method is inspired by a statistical-mechanical approach to systems governed by dynamics with largely-separated time scales and is based on three key concepts: conjugate pairs of variables, dimensionless integration measures with coarse-graining factors and partial maximization of the joint entropy. The method enables one to calculate a prior purely from a likelihood in a simple way. It is shown in particular how it not only yields Jeffreys's rules but also reveals new structures hidden behind them.

Bayesian statistics plays prominent roles in diverse areas in sciences and technologies. It has experienced a number of vicissitudes in the past, but today it is regarded as an efficient theory for inference and decision making. (For a nice general reading, see [1].)

In Bayesian inference of a quantity from data with uncertainty, selection of a prior probability is a central issue, and accordingly such a problem has in fact been repeatedly discussed in the literature. Since Bayesian theory has its history of over 250 years, it is not possible to overview them, here. (Instead, we quote [2-5].) However, among others, the discussions of Jeffreys [6] and Jaynes [7] keep outstanding positions for our interest in the sense that they are based on the invariance principle for the information-theoretic quantities. Jeffreys has investigated the transformation property of a matrix, which is termed the Fisher information today and is actually a Riemannian metric induced in a space of parametrized probability distributions. Jaynes has introduced an invariant measure inside the logarithm in the definition of the entropy and then imposed the reparametrization invariance on a prior as a maximum-entropy distribution to eliminate an ambiguity in determining the measure. Both of them enable one to select specific priors for parameter estimations. It is however noted that two priors for the same problem determined by the methods of Jeffreys and Jaynes do not coincide with each other, in general, and it is interesting to see that actually it is Jaynes's that can reproduce Jeffreys's original priors [8].

In this paper, we present a new information-theoretic approach to the problem of selecting priors in Bayesian statistics, which is refereed to here as the *conditional* *maximum-entropy method* (abbreviated as C-MaxEnt). This method is inspired by statistical mechanics of systems governed by hierarchical dynamics with largely-separated time scales, which can be implemented via the conditional concepts [9,10] including conditional probabilities (i.e., likelihoods) and conditional entropies. In accordance with the statistical-mechanical procedure, we do not introduce any invariant measure in the definition of the entropy, in contrast to Jaynes's proposal [7,11]. As a result, C-MaxEnt enables us to obtain in a very simple way a prior probability distribution that is purely given in terms of a likelihood. We apply C-MaxEnt to several typical examples to see how Jeffreys's rules are naturally derived. And, more importantly, we show that this method can reveal new structures hidden behind the rules.

We start our discussion by presenting our idea of C-MaxEnt. This method consists of three key ingredients:

    I)    find *conjugate pairs of variables*;

    II)   define a dimensionless integration measure with coarse-graining factors;

    III)  maximize the joint entropy only with respect to a prior,

where "conjugate pair of variables" in I) should not be confused with "conjugate prior". Below, we explain I)-III) in detail, in order to see how this idea is intimately related to statistical mechanics. I) and II) may actually be linked together, as will immediately be seen.

To illustrate the idea as simple as possible, let us consider a couple of continuous random variables, $X$ and $\Theta$, whose realizations are $x$ and $\theta$, respectively. The joint probability distribution reads $P(x,\theta)$. To specify the normalization condition, it is necessary to define an integration measure.

In statistical mechanics, the integration measure is given by a dimensionless one, $\prod_{i=1}^{N}(d^3\mathbf{Q}_i\, d^3\mathbf{P}_i/h^3)$ (with possible Gibbs's factor), where $N$ is the number of particles, $\mathbf{Q}_i$ ($\mathbf{P}_i$) is the canonical coordinate (momentum) of the $i$-th particle and $h$ is the Planck constant. A point is that $h$ is a coarse-graining factor [12] and is of crucial importance in order for the entropy to exist since the entropy is definable only for countable sets (Boltzmann's principle). It should be noted that $h$ does not explicitly appear at the level of thermodynamics (i.e., in the equation of state, the specific heat and so on) and is absorbed into the definition of the thermal wavelength, which is an experimentally observable quantity.

Analogously to statistical mechanics, we wish to introduce the concept of conjugate pairs of variables. We shall see that this concept is important for eliminating invariant measures from the definition of the entropy. A clue is that the above-mentioned integration measure in statistical mechanics is invariant under time evolution (or, more generically, canonical transformations) since $\mathbf{Q}_i$ and $\mathbf{P}_i$ ($i=1,2,...,N$) are canonically-conjugate variables, whereas $X$ and $\Theta$ in our case are not canonical variables, in general. To introduce the concept of a *conjugate pair of variables*, first we factorize the joint probability distribution as follows:

$$P(x,\theta) = p(x|\theta) f(\theta), \qquad (1)$$

where $p(x|\theta)$ is a conditional probability distribution to be identified with a likelihood and $f(\theta)$ is a prior. Our proposal is then as follows: if the conditional variance, $(\Delta X)^2$, with respect to an integration measure to be defined behaves as $(\Delta X)^2 \propto 1/\theta^2$, then $(X, \Theta)$ is said to be a conjugate pair. (Here, clearly the second conditional moment is assumed to exist. Later, we shall discuss the case when the second moment does not exist.) Accordingly, it is natural to define the dimensionless integration measure as follows:

$$d\Gamma = \left(\frac{dx}{k}\right)\left(\frac{d\theta}{l}\right). \tag{2}$$

Here, $k$ and $l$ are coarse-graining factors with the dimensions of $X$ and $\Theta$, respectively. Although it turns out that these factors do not explicitly appear in priors obtained, still they are indispensable in the definition of the entropy. The normalization conditions read: $\int d\Gamma P(x,\theta) = 1$, $\int (dx/k) p(x|\theta) = 1$ and $\int (d\theta/l) f(\theta) = 1$. These procedures may illustrate I) and II).

In the above, we have mentioned that although the Planck constant is necessary even in classical statistical mechanics, it does not explicitly appear at the level of thermodynamics (of classical systems). As we shall see below, this situation is similar here. Although the coarse-graining factors, e.g. $k$ and $l$ in eq. (2), should be introduced for systems described by continuous variables, they do not explicitly appear in results.

Now, we explain III). As already mentioned, in statistical mechanics, no invariant measures are introduced inside the logarithm in the definition of the entropy. Therefore, as the joint entropy, we employ the differential entropy

$$S[X,\Theta] = -\int d\Gamma\, P(x,\theta) \ln P(x,\theta). \tag{3}$$

Substitution of eq. (1) into eq. (3) yields

$$S[X,\Theta] = S[X|\Theta] + S[\Theta]. \tag{4}$$

Here, the two quantities on the right-hand side are the conditional and marginal entropies respectively given by

$$S[X|\Theta] = \int \frac{d\theta}{l} f(\theta)\, S[X|\theta], \qquad S[\Theta] = -\int \frac{d\theta}{l} f(\theta) \ln f(\theta), \tag{5}$$

with $S[X|\theta)$ being

$$S[X|\theta) = -\int \frac{dx}{k} p(x|\theta) \ln p(x|\theta), \tag{6}$$

which will play a primary role in the subsequent discussion. (The notation, $S[X|\theta)$, indicates that it is a function of $\theta$ but not of $x$.)

The factorization in eq. (1), too, has an analogy in statistical mechanics, especially of a system governed by dynamics with largely-separated time scales. Suppose that $X$ is a fast variable, whereas $\Theta$ is a slow one. The total dynamics has a hierarchical structure characterized by this time-scale separation. In this situation, it makes sense to consider the conditional probability distribution of $X$, given a value of $\Theta$, but not vice versa. Such a distribution is precisely $p(x|\theta)$ in eq. (1), and $f(\theta)$ describes its slow modulation [10]. In accordance with the standard treatment of fast and slow degrees of freedom in physics, the fast variable is integrated/eliminated before the slow one. And, the dynamics of the fast variable does not affect that of the slow one. This naturally leads to the partial maximization of the joint entropy only with respect to $f(\theta)$. So, III) is formulated as follows:

$$\delta_f \left\{ S[X, \Theta] - \alpha \left( \int \frac{d\theta}{l} f(\theta) - 1 \right) \right\} = 0, \tag{7}$$

where $\delta_f$ denotes the variation with respect to $f(\theta)$, and $\alpha$ is Lagrange's multiplier associated with the constraint on the normalization condition. From eq. (4), we obtain the prior

$$f(\theta) \propto \exp\{S[X \mid \theta]\} \tag{8}$$

with $S[X \mid \theta]$ given in eq. (6). This is, in fact, independent of $x$, as desired.

Thus, C-MaxEnt allows us to calculate a prior from a likelihood in a simple and self-contained way.

Before proceeding, we wish to make the following four remarks. Firstly, to derive eq. (8), we have imposed a constraint only on the normalization condition. Therefore, $f(\theta)$ in eq. (8) is regarded to be the so-called non-informative prior. If we are given additional knowledge about, e.g., the expectation value $\bar{E}$ of a certain quantity $E(\Theta)$, $\int (d\theta / l) E(\theta) f(\theta) = \bar{E}$, then we have a modified form of eq. (8): $f(\theta) \propto \exp\{S[X \mid \theta] - \beta E(\theta)\}$, where $\beta$ is Lagrange's multiplier associated with the constraint on the expectation value. (Clearly, more constraints can be imposed, in general, as long as they are consistent with each other.) However, here we do not do so and focus our attention on the nature of eq. (8). Secondly, the procedure described in eq. (7) is opposite to the one discussed in [13,14], where the joint entropy is maximized with respect to a conditional probability distribution. Thirdly, the partial maximization in eq. (7) has been discussed in [9,15,16]. However, in these earlier works, the points in I) and II) have never been addressed, and accordingly the resulting priors given there may be different from what obtained from C-MaxEnt, in general (see the next discussion). Fourthly, we emphasize that other entropic approaches presented in [17-20] use Jaynes's invariant measure and are also conceptually different from C-MaxEnt again due to lack of the keys I) and II).

Now, we discuss several typical applications of C-MaxEnt.

As the first example, let us consider the case when $\Theta$ is an *inverse* scale variable, that is,

$$p(x|\theta) = \theta \tilde{p}(\theta x), \tag{9}$$

where $\tilde{p}(x)$ satisfies the normalization condition, $\int (dx/k)\, \tilde{p}(x) = 1$. Henceforth, the notations, $\tilde{p}$, will commonly be used for different problems, but this will not cause any confusion. In turn, the scale variable is given by

$$\sigma = \frac{1}{\theta}. \tag{10}$$

In this case, $\Theta$ has to be a positive random variable. If the first two conditional moments exist, then we have $<X> \equiv \int (dx/k) \, x \, p(x|\theta) = c_1/\theta$ and $\langle X^2 \rangle = c_2/\theta^2$, where $c_1$ and $c_2$ are constants respectively given by $c_1 = \int (dx/k) \, x \, \tilde{p}(x)$ and $c_2 = \int (dx/k) \, x^2 \tilde{p}(x)$, leading to the conditional variance

$$(\Delta X)^2 = \frac{c_2 - c_1^2}{\theta^2}. \tag{11}$$

This implies that the conjugate pair of variables is $(X, \Theta)$, and eq. (2) may be employed as an integration measure. Substitution of eq. (9) into eq. (6) gives rise to

$$S[X|\theta] = -\ln\theta + C_1, \tag{12}$$

where $C_1 = -\int (dx/k) \, \tilde{p}(x) \ln \tilde{p}(x)$ is a constant. Therefore, from eq. (8), we have the following prior:

$$f(\theta) \, d\theta \propto \frac{d\theta}{\theta}. \tag{13}$$

Using $\sigma$ in eq. (10), we alternatively have

$$\frac{d\sigma}{\sigma}, \tag{14}$$

which is precisely Jeffreys's original rule for the scale [8]. Note that this cannot be obtained, if a pair $(X, \Theta^{-1})$ is employed, for example: the use of $(X, \Theta^{-1})$ gives rise to the rule of the form $\sigma\, d\sigma$, which is not favorable because of its monotonic increase with respect to the scale variable. This highlights how I) is of crucial importance.

Regarding a prior for $\sigma\,(=1/\theta)$, the scaling in eq. (9) is actually sufficient for identifying $(X,\Theta)$ with the conjugate pair of variables, even if $p(x|\theta)$ does not have the conditional second moment. An example may be supplied by the Lévy distribution: $p(x|\theta) = k\, L_\alpha(x;\theta) = (k/2\pi)\int_{-\infty}^{\infty} d\kappa\, \exp\left(-i\kappa x - |\kappa|^\alpha/\theta\right)$, where $0 < \alpha < 2$. The factor $k$ in front of $L_\alpha(x;\theta)$ is introduced to preserve the normalization condition, $\int (dx/k)\, p(x|\theta) = 1$. This distribution decays as a power law: $L_\alpha(x;\theta) \sim |x|^{-1-\alpha}$ ($x \to \pm\infty$) and therefore does not have the second moment. Clearly, this $p(x|\theta)$ satisfies the scaling of the form: $p(x|\theta) = \theta^{1/\alpha}\,\tilde{p}(\theta^{1/\alpha} x)$, which indicates that a conjugate pair is $(X,\Theta^{1/\alpha})$ and accordingly the integration measure is $d\Gamma = (dx/k)(d\theta^{1/\alpha}/l')$ (with $l'$ being different from $l$ in eq. (2), in general). Then, C-MaxEnt again yields eq. (14), as it should do.

The second example is concerned with the case when some variables do not have their conjugate partners. To discuss it in a simple way, let us combine the scale with the location. Thus, we take three random variables $X$, $M$ and $\Theta$ with their realizations $x$, $\mu$ and $\theta$, respectively. Here, $M$ is the location variable. Both $x$ and $\mu$ are arbitrary real, whereas $\theta$ is positive like before. We write the joint probability distribution as follows:

$$P(x,\mu,\theta) = p(x|\mu,\theta)\, f(\mu,\theta)\,. \tag{15}$$

We sample from the conditional probability distribution satisfying

$$p(x|\mu,\theta) = \theta\, \tilde{p}(\theta(x-\mu)). \tag{16}$$

As emphasized in [7], this is one of the most important cases in practice. A conjugate pair is $(X, \Theta)$. On the other hand, $M$ does not have its conjugate partner. We introduce three coarse-graining factors $k$, $l$ and $m$, and take the integration measure

$$d\Gamma = \left(\frac{dx}{k}\right)\left(\frac{d\mu}{m}\right)\left(\frac{d\theta}{l}\right). \tag{17}$$

Accordingly, $\tilde{p}(x)$ in eq. (16) is normalized as follows: $\int (dx/k)\, \tilde{p}(x) = 1$. Now, C-MaxEnt gives rise to the prior

$$f(\mu,\theta) \propto \exp\{S[X|\mu,\theta]\}. \tag{18}$$

From eq. (16), the function $S[X|\theta)$ is calculated to be

$$S[X|\mu,\theta] = -\ln\theta + C_2, \tag{19}$$

where $C_2 = -\int (dx/k)\, \tilde{p}(x) \ln \tilde{p}(x)$ is a constant. Therefore, we have

$$f(\mu,\theta)\, d\mu\, d\theta \propto \frac{d\mu\, d\theta}{\theta}, \tag{20}$$

or, in terms of the scale variable in eq. (10),

$$\frac{d\mu\, d\sigma}{\sigma}. \qquad (21)$$

Again, this is the original Jeffreys rule for the scale and location [8].

In general, in order to define an appropriate integration measure, it is necessary to *identify conjugate pairs of variables as many as possible*.

So far, we have only seen how C-MaxEnt can reproduce known rules of Jeffreys's for non-informative priors. However, actually C-MaxEnt does tell us more than just that. Therefore, finally we discuss the problem concerning the Poisson distribution

$$p(n|\theta) = e^{-\theta}\frac{\theta^n}{n!}. \qquad (22)$$

In this case, the realization of $X$ is $n = 0, 1, 2, \ldots$ and $\theta$, the realization of $\Theta$, is positive. Clearly, both of them are free from physical dimensions. Note that $\Theta$ may not be a scale variable, here. As well-known, the conditional variance is given by $(\Delta X)^2 = \theta$. This indicates that $(X, \Theta^{-1/2})$ is a conjugate pair of variables. The corresponding integration measure reads

$$\int d\Gamma\, (\cdots) = \sum_n \int \frac{d\theta}{l_0\, \theta^{3/2}}(\cdots), \qquad (23)$$

where $l_0$ is a dimensionless coarse-graining factor associated with $\theta$ (no coarse-graining factor is needed for $n$ because of its discreteness). The quantity, $S[X|\theta] = -\sum_n p(n|\theta)\ln p(n|\theta)$, is evaluated as follows:

$$S^{\text{Poisson}}[X|\theta] = \theta - \theta \ln\theta + e^{-\theta} \sum_{n=2}^{\infty} \frac{\theta^n}{n!} \ln(n!) \,. \tag{24}$$

It is unlikely that the summation in the last term on the right-hand side can explicitly be performed. So, we have numerically evaluated the prior

$$\frac{f(\theta)}{\theta^{3/2}} d\theta \propto \frac{d\theta}{\theta^{3/2}} \exp\{S^{\text{Poisson}}[X|\theta]\}, \tag{25}$$

where the factor $1/\theta^{3/2}$ comes from the integration measure in eq. (23). The result is given in fig. 1. There, one observes a remarkable fact: eq. (25) well reproduces Jeffreys's original rule for the Poisson parameter [8]

$$\frac{f(\theta)}{\theta^{3/2}} d\theta \propto \frac{d\theta}{\theta} \tag{26}$$

up to $\theta \cong 100$, then indicates the existence of a crossover and finally decays very rapidly, implying that the rule in eq. (26) is applicable for $\theta$ less than about 100. This is a nontrivial result revealed by C-MaxEnt and gives a new insight into the problem. The modification of Jeffreys's rule shown in fig. 1 is quite suggestive. $\theta$ in the Poisson distribution is not a factor related to the scale variable. Therefore, there is no *a priori* reason to assume that the rule for the inverse scale variable in eq. (13) holds also for the prior of the Poisson parameter, here. Recalling the fact that both the expectation value and the variance of $X$ are $\theta$ in the Poisson distribution, $\theta \cong 100$ is understood to be very large. The rapid decay after $\theta \cong 100$ suppresses such large fluctuations.

In conclusion, we have formulated C-MaxEnt for selecting prior probability distributions in Bayesian statistics, being inspired by a statistical-mechanical approach to systems governed by hierarchical dynamics with largely-separated time scales. We have set up three key concepts: conjugate pairs of variables, dimensionless integration measures with coarse-graining factors and partial maximization of the joint entropy, as in (I)-(III). We have seen how this method allows us to calculate a prior purely from a likelihood in a simple way. As applications, we have shown how C-MaxEnt not only reproduces Jeffreys's rules but also reveals hidden structures behind them.

\* \* \*

The author would like to thank RAMANDEEP S. JOHAL, STEVE PRESSÉ, NORIKAZU SUZUKI and RAVI C. VENKATESAN for discussions. This work was supported in part by a Grant-in-Aid for Scientific Research from the Japan Society for the Promotion of Science.

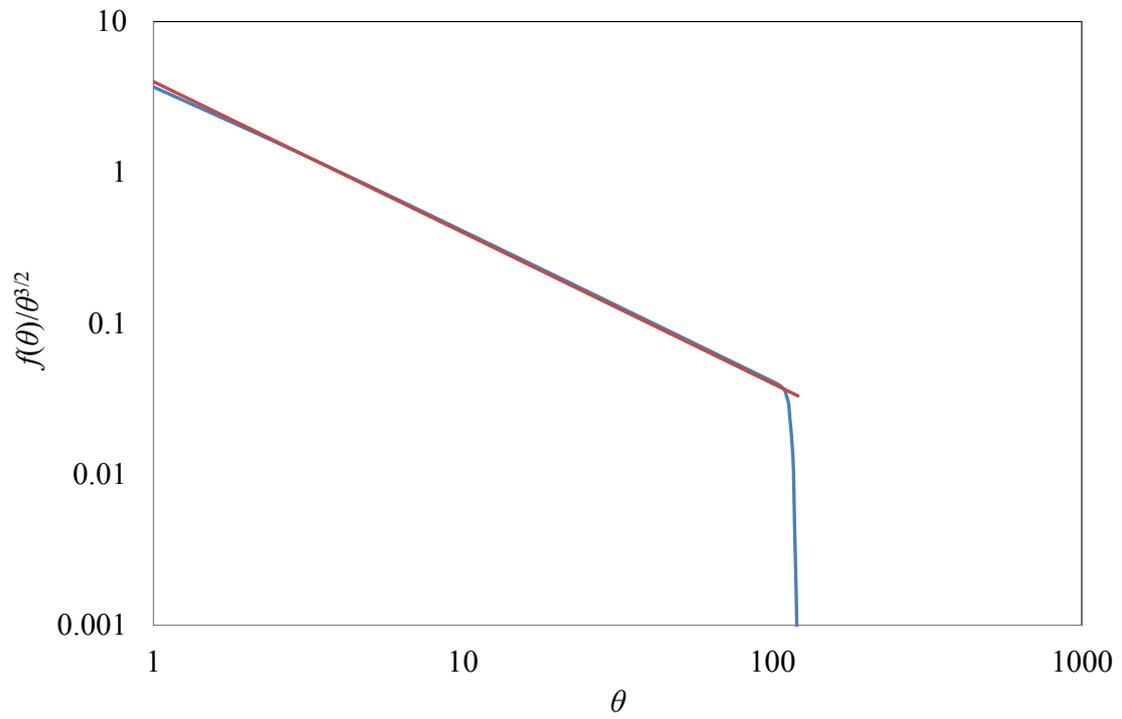

**Fig. 1** The log-log plot of unnormalized $f(\theta)/\theta^{3/2}$ in Eq. (25) with respect to $\theta$. The read line is the rule in Eq. (26). All quantities are dimensionless.